\documentclass[preprint]{aastex631}

\usepackage{amsmath,esint}
\usepackage{subfigure}
\usepackage{appendix}
\usepackage{bm}
\graphicspath{{./}{figures/}}

\begin{document}

\title{Revisiting the {\em Chandra} Observation on the Region of PSR~J1809-1917: Indication of an X-ray Halo and Implication for the Origin of HESS~J1809-193}
\author[0000-0002-0786-7307]{Chao-Ming Li}
\affiliation{School of Astronomy and Space Science, Nanjing University, Nanjing 210023, China}
\author{Chong Ge}
\affiliation{Department of Astronomy, Xiamen University, Xiamen, Fujian 361005, China}
\author[0000-0003-1576-0961]{Ruo-Yu Liu}
\affiliation{School of Astronomy and Space Science, Nanjing University, Nanjing 210023, China}
 \affiliation{Key laboratory of Modern Astronomy and Astrophysics (Nanjing University), Ministry of Education, Nanjing 210023, China}
\correspondingauthor{Ruo-Yu Liu}
\email{ryliu@nju.edu.cn}

\begin{abstract}
HESS~J1809-193 is an extended TeV $\gamma$-ray source and the origin of its $\gamma$-ray emission remains ambiguous. Pulsar wind nebula (PWN) of PSR J1809-1917 laying inside the extended $\gamma$-ray emission is a possible candidate. Powered by the central pulsar, ultrarelativistic electrons in PWN can produce radio to X-ray emission through synchrotron and $\gamma$-ray emission by inverse Compton (IC) scattering. To check whether this PWN is the counterpart of HESS~J1809-193, we analyzed {\em Chandra} X-ray radial intensity profile and the spectral index profile of this PWN. We then adopt a one-zone isotropic diffusion model to fit the keV and the TeV data. We find diffuse nonthermal X-ray emission extending beyond PWN, which is likely an X-ray halo radiated by escaping electron/positron pairs from the PWN. A relatively strong magnetic field of $\sim 20\,\mu$G is required to explain the spatial evolution of the X-ray spectrum (i.e., the significant softening of the spectrum with increasing distance from the pulsar), which, however, would suppress the IC radiation of pairs. Our result implies that a hadronic component may be needed to explain HESS~J1809-193.
\end{abstract}

\section{Introduction}
A pulsar wind nebula (PWN) is believed to be a bubble of energetic electron/positron pairs, radiating broadband nonthermal emission from radio band to multi-TeV or even PeV $\gamma$-ray band. These energetic pairs are suggested to be accelerated at the termination shock driven by the interaction between the relativistic pulsar wind and ambient medium. From the perspective of energy budget, those energetic pairs are essentially powered by the rotational energy of the pulsar \citep{gaensler_evolution_2006}. Therefore, relatively young pulsars with high spindown luminosity are naturally efficient particle accelerators. Indeed, nonthermal X-ray radiations and TeV $\gamma$-ray radiations have been observed from many young PWN, which likely arise from, respectively, the synchrotron radiation of relativistic electrons (hereafter we do not distinguish positrons from electrons, unless otherwise specified) in the magnetic field and inverse Compton (IC) scattering against background photon fields. While the energy of X-ray emitting electrons are subject to the uncertainty of magnetic field strength in the PWN, the energy of multi-TeV-emitting electrons are generally found to be of at least $\sim 10-100\,$TeV since the IC target radiation field are more or less known (i.e, interstellar infrared radiation field and the cosmic microwave background).
Indeed, among all the TeV $\gamma$-ray sources detected by the High Energy Spectroscopic System (HESS) in the Galactic plane survey, identified PWN or PWN candidates constitute the largest population \citep{2018A&A...612A...1H}. We also see associations of energetic pulsars with many ultrahigh-energy $\gamma$-ray (with photon energy above 100\,TeV) sources detected by the Large High Altitude Air Shower Observatory (LHAASO). Therefore, it would be natural to expect TeV emissions from PWN of energetic pulsars.

HESS J1809-193 is an extended TeV $\gamma$-ray source without a firmly identified physical association. It was first discovered by HESS \citep{aharonian_discovery_2007}, with the spectrum continuing to $\sim 10$\,TeV. The High Altitude Water Cherenkov Observatory (HAWC) reported an extended source at the same position above 56\,TeV (i.e., eHWC~J1809-193, \citealt{abeysekara_multiple_2020}) and the spectrum probably extends up to 177\,TeV (i.e, 3HWC~J1809-190, \citealt{albert_3hwc_2020}). This source was initially suggested to be the PWN powered by PSR~J1809-1917, which is an energetic pulsar of spindown luminosity $L=1.8\times 10^{36}\, \rm erg \, s^{-1}$ with a characteristic age of $\tau_{c}=50\,$kyr and distance of about 3.5 kpc via the IC radiation of accelerated electrons in the PWN \citep{aharonian_discovery_2007}. This speculation was supported by the discovery of the bright X-ray PWN of the pulsar \citep{kargaltsev_x-ray_2007}. \citet{klingler_variable_2018} divided the PWN into the compact nebula and extended nebula according to their intensity but no spectral differences were detected. Their further investigation \citep{klingler_chandra_2020} suggested that no other source in this field may reproduce the TeV flux of HESS J1809-193 other than PWN of PSR J1809-1917. However, the TeV nebula observed by HESS has a 70\% containment radius of $0.64^\circ$, which is far more extended than the size of the X-ray PWN which is about 2$'$. 

\citet{di_mauro_evidences_2020} interpreted the extended TeV emission as the IC radiation of escaping electrons from the PWN. Assuming a suppressed particle diffusion in the ISM around the PWN, they showed that the surface brightness profile of the TeV gamma-ray emission can be reproduced, which is similar to the scenario of the so-called pulsar halo \citep{2022NatAs...6..199L, liu_physics_2022,2022FrASS...922100F}. Pulsar halos are generally believed to be generated around middle-aged pulsars older than 100\,kyr, when the pulsars have left associated SNR and are traversing ISM. At that stage, accelerated electrons can largely escape the PWN and diffuse in the ISM. Although the characteristic age of PSR~J1809-1917 is about 50\,kyr, a fraction of accelerated electrons may already escape. Indeed, escape of electrons from relatively young PWNe has been hinted in previous literature, such as for Vela X \citep{hinton_escape_2011} and HESS~J1825-137 \citep{liu_unusually_2020}. 

However, the consensus of the IC origin of HESS~J1809-193 is yet to be reached. JVLA observation performed by \cite{castelletti_unveiling_2016} found no radio counterpart to the observed X-ray emission supposed to be a PWN but discovered some molecular clouds instead. They concluded that HESS J1809-193 originated from the collisions of ions in the SNR G11.0-0.0 with the molecular cloud by a hadronic mechanism. Fermi-LAT observation in GeV performed by \cite{araya_gev_2018} supported the hadronic scenario and they also declared that a mixed scenario involving IC emission and hadronic emission could not be discarded. 

It has been indicated \citep{liu_constraining_2019, Liben22} that electrons that radiate multi-TeV emission will produce keV emission in the typical interstellar magnetic field (i.e, a few $\mu$G). If HESS J1809-193 originates from the IC radiation of escaping electrons from the PWN, we should expect diffuse synchrotron X-ray radiation surrounding the X-ray PWN of PSR~J1809-1917, permeating throughout the entire region of the TeV emission with a similar morphology. We can get a hint of this scenario from the observation of Suzaku in this region \citep{Anada2010, Bamba2010}, where diffuse X-ray emission is found to extend beyond the PWN toward south. In order to further test this scenario, we attempt to search for the diffuse X-ray emission surrounding the bright, compact X-ray PWN, and check whether it is consistent with the IC interpretation of HESS~J1809-193. The rest of the paper is organized as follows. In Section 2 we show our analysis of 14 {\em Chandra}
ACIS observations on the PSR J1809-1917 field. In Section 3 we present the results of the X-ray data analysis and physical implications for the origin of the TeV emission. We discuss the origin of the TeV emission and summarize our conclusions in Section 4. 

\section{Observation and Data Reduction} \label{sec:displaymath}
\begin{deluxetable}{ccccc}
\tablecaption{{\em Chandra} ACIS-I observations of PSR J1809-1917\label{table:observation}}
\tablewidth{0pt}
\tablehead{
\colhead{Obsid} & \colhead{Inst} & \colhead{Time} &
\colhead{Clean time} & \colhead{Obsdate} 
}
\decimalcolnumbers
\startdata
20327  & ACIS-I & 15.3 & 15.3 & 2018-02-08 \\ 
20328  & ACIS-I & 33.6 & 33.2 & 2018-04-02 \\ 
20329  & ACIS-I & 29.7 & 29.5 & 2018-05-25 \\ 
20330  & ACIS-I & 32.6 & 32.6 & 2018-07-18 \\ 
20331  & ACIS-I & 39.5 & 39.5 & 2018-08-30 \\ 
20332  & ACIS-I & 37.6 & 37.5 & 2018-11-03 \\ 
20962  & ACIS-I & 20.9 & 20.8 & 2018-02-09 \\ 
20963  & ACIS-I & 16.3 & 16.2 & 2018-02-12 \\ 
20967  & ACIS-I & 15.1 & 15.0 & 2018-02-10 \\ 
21067  & ACIS-I & 30.6 & 30.5 & 2018-04-03 \\ 
21098  & ACIS-I & 39.4 & 39.2 & 2018-05-27 \\ 
21126  & ACIS-I & 40.0 & 40.0 & 2018-07-19 \\ 
21724  & ACIS-I & 25.7 & 25.7 & 2018-09-01 \\ 
21874  & ACIS-I & 28.7 & 28.6 & 2018-11-04 \\ \enddata
\tablecomments{(1) {\em Chandra} observation ID. (2) Observation instrument. (3) Total exposure time (unit: ks). (4) Cleaned exposure time after filter (unit: ks). (5) Observation date.}
\end{deluxetable}
We chose 14 {\em Chandra} Advanced CCD Imaging Spectrometer (ACIS)-I observations (405.1\,ks exposure time in total) of PSR J1809-1917 in 2018. These observations were all operated in timed exposure Full Frame Very Faint mode, for which the time resolution is 3.24 s. Details of these observations are listed in Table \ref{table:observation}. These data were dealt with {\em Chandra} Interactive Analysis of Observations software package (CIAO) version 4.14 as well as Calibration Database (CALDB) version 4.9.8\footnote{\url{https://cxc.harvard.edu/ciao/index.html}}. After reprocessing the original datasets, we merged the $0.5-7$\,keV images with {\sc merge}\_{\sc obs} by setting the parameter `bands = broad'. Then we used {\sc wavdetect} for the merged event to detect the X-ray point sources ($> 5\sigma$) which would be masked in later analysis. We also manually chose 11 other bright regions as point sources to be masked. There are 319 point sources masked in total.  

If the IC interpretation of HESS~J1809-193 is right, there would exist a very diffuse and extended X-ray halo in the entire field of view (FoV) of {\em Chandra}. Thus, we have to analyze and subtract the background of {\em Chandra} observation carefully. {\em Chandra} ACIS-I background consists of four components \citep{hickox_absolute_2006}: readout artifact, real cosmic X-ray background, flaring detector background, and quiescent particle background. Readout artifact due to the finite readout time of ACIS detectors has little effect thus we ignored it. The real cosmic X-ray background mainly contributes photons below $\sim 1$ keV and was excluded in the  spectra fitting. We removed the flaring detector background ($> 3\sigma$) by analyzing the lightcurve of $2.3-7.3$\,keV band which is the most sensitive to flares. Then we derived the clean events with a total exposure of 403\,ks. To subtract the quiescent particle background, we used $9.5-12$\,keV photons of the clean events to rescale the blank-sky events\footnote{\url{https://cxc.harvard.edu/ciao/threads/acisbackground/}}. The effective area of ACIS instrument in this energy range is so small that nearly all signal comes from quiescent particle background. Then we merged the clean events, removed the point sources, filled the holes, and subtracted the blank-sky files to get a clean image. This image was further divided by the exposure map to get the real diffuse emission of PWN\footnote{\url{https://cxc.harvard.edu/ciao/threads/diffuse_emission/}}.

To measure the intensity profile and the spectral profile of the X-ray PWN and the possibly existed X-ray halo, we divide the region into several sectors based on the distance to the pulsar and extract spectra of each individual region, as shown in Fig.~\ref{fig:realpwn}. \citet{klingler_chandra_2020} suggested that a kinetic jet appears in the southeast part of the PWN. We therefore divide the J1809 field into region {\sl a} (green) and region {\sl b} (white) to separate this possible structure from the rest part of the extended emission. Then we extract the X-ray properties of these regions with the clean events and the corresponding blank-sky files for each observation. We use clean events as source spectra while blank-sky files as background spectra. 
We manually rescale the background spectrum with the count rate of $9.5-12$\,keV photons to subtract background spectrum correctly. Finally, we jointly fit the spectra and calculate flux in {\sc xspec}. 

We also used {\em Chandra} stowed background to cross-check with our spectra analysis using blank-sky background. Stowed background means ACIS runs in the normal imaging very faint mode but further away from the external calibration source to avoid the residual line illumination\footnote{\url{https://cxc.harvard.edu/contrib/maxim/stowed/}}. Therefore it only consists of quiescent particle background and can be used just like blank-sky files. The only difference is additional spectra components have to be considered in spectra fitting.

\section{Results} 
\subsection{PWN Morphology and Spectra Analysis }\label{sec:3.1}

\begin{figure}[htbp]
    \centering
    \includegraphics[width=0.5\columnwidth]{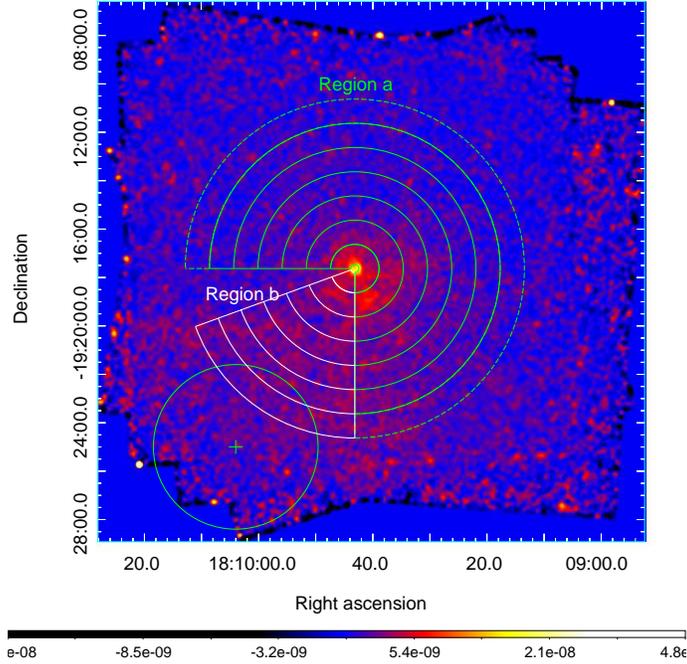}
    \caption{Exposure-corrected image of diffuse emission of the PWN on the merged ACIS-I events with quiescent particle background subtracted, binned by a factor of 2 and smoothed with a 10-pixel (r=10$''$) Gaussian kernel. Point sources are removed. PSR J1809-1917 lies in the center of the annuli. Green (namely region {\sl a}) and white (namely region {\sl b}) annular sectors are selected to extract spectra separately. Dashed green annular sector was chosen as local background region. The green circle marks the shell of SNR G11.0-0.0 in the radio band and the green cross is the center of this SNR.}
    \label{fig:realpwn}
\end{figure}

The smoothed, exposure-corrected image of the extended emission after the subtraction of the quiescent particle background is displayed in Fig. \ref{fig:realpwn}. We find a bright core in the center with an extended structure expanding towards southeast, which is generally consistent with  \citet{klingler_chandra_2020}. It may be worth noting that we do not spot a jet-like structure present in the southeast region as reported by \citet{klingler_chandra_2020}. The difference might be caused by that they do not subtract the quiescent particle background. Instead, the extended structure in our image appears a sector-like structure with the apex being consistent with the pulsar's position. We speculate that it may be the tail of PWN, where accelerated particles are advecting away from the PWN. If so, the pulsar may be moving northwest, which is consistent with the direction of the pulsar's proper motion suggested by \cite{klingler_chandra_2020}, i.e., $\mu_{\alpha} = -1\pm9 \,  {\rm mas \,  yr}^{-1}$ and $\mu_{\delta} = 24\pm11 \, {\rm mas \, yr}^{-1}$. Alternatively, we note that the SNR G11.0-0.0 presents in the southeast part of the PWN, marked as the green circle in Fig.~\ref{fig:realpwn}. According to \cite{brose_morphology_2021}, the most energetic particles tend to reside outside SNR, which might also contribute the X-ray emission in this region. But a shell-like morphology in X-ray is not seen, making it unlikely to be an SNR halo.  

\begin{figure}[htbp]
    \centering
    \includegraphics[width=0.5\columnwidth]{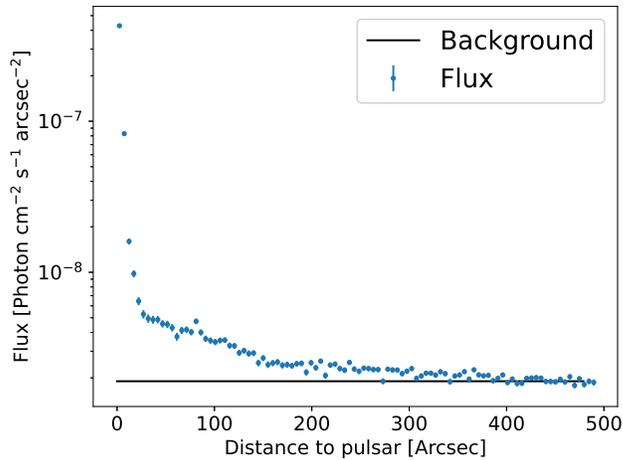}
    \caption{Radial profile of flux with a bin size of 5$''$ on the point sources excluded, quiescent particle background subtracted and exposure-corrected image. The solid black line is the estimated cosmic X-ray background by the flux farther than 400$''$.}
    \label{fig:rp}
\end{figure}

As the first trial, we extracted the radial profile of X-ray photons around PSR~J1809-1917 based on the reduced image. As shown in Fig.~\ref{fig:rp}, the obtained profile shows a steep decline in flux with increasing distance from the pulsar within a distance of 30$''$. Afterwards, the decline becomes shallow, with some fluctuations that could be caused by the small bin size used. Overall, this profile may suggest the presence of an X-ray halo extending beyond the PWN, as evidenced by the continuous decrease in the profile up to a distance of about 400$''$. 

To test the IC interpretation of HESS~J1809-193, we will avoid using the emission from the `tail' in the southeast but focus on the possible X-ray emission in the other three quarters. Since the entire region could be overwhelmed by the escaping electrons, we will employ two different strategies for the astrophysical background subtraction in the following two subsections.  


\subsubsection{Spectral Analysis with Local Background}\label{sec:3.1.1}

We have divided the field into two regions (Fig.~\ref{fig:realpwn}): one is in the southeast covering the extended `tail' (namely region {\sl b}) and the other covering three-quarters of the field (namely region {\sl a}). We used annular sector regions to analyze the intensity and spectral distribution of these two regions. In this subsection, the outermost annular sector of region {\sl a} is chosen as the local background to fit the residual of the cosmic X-ray background. We used the double background subtraction \citep{geng_chandra_2022} and jointly fit the spectra in {\sc xspec}  with {\sc apec+tbabs*powerlaw+tbabs*powerlaw}. The two {\sc tbabs} are different, one is for the total line of sight and the other is for the location near the PWN inside the Galaxy. {\sc apec+tbabs*powerlaw} aims to fit the residual of cosmic X-ray background and we set the Hydrogen column density $N_{\rm H}$ to be $1.60 \times 10^{22} $ ${\rm cm}^{-3}$, calculated from X-Ray Background Tool\footnote{\url{https://heasarc.gsfc.nasa.gov/cgi-bin/Tools/xraybg/xraybg.pl}}. The second {\sc tbabs*powerlaw} aims to fit the PWN \& SNR diffuse emission and we set the $N_{\rm H}$ to be $0.7 \times 10^{22}{\rm\ cm}^{-3}$, the same as \cite{klingler_chandra_2020}.

We ignored the energy band lower than 0.7\,keV where {\em Chandra} ACIS detector is not sensitive enough any longer due to the accumulated contamination on the ACIS optical blocking filters. We also ignored the energy band higher than 7\,keV where the detector background dramatically rises. The solar abundance was set to {\sc aspl} which adopted the value from \cite{annurev.astro.46.060407.145222}.  The spectra of the residual cosmic X-ray background were linked for the PWN and local background regions and their normalization is rescaled by the solid angles of regions.

For stowed background analysis, we also used the double background subtraction (\citealt{Ge2021}) and jointly fit the spectra in {\sc xspec}. The spectra model is {\sc apec+tbabs*(apec + powerlaw) + tbabs*powerlaw}. Since stowed background only consists of quiescent particle background, we used {\sc apec+tbabs*(apec+powerlaw)} to fit X-ray sky background and {\sc tbabs*powerlaw} to fit PWN \& SNR emission. 

\begin{deluxetable*}{CCCCCCC}
\tabletypesize{\scriptsize}
\tablecaption{ Results of spectral fitting with local background\label{table:results}}
\tablewidth{0pt}
\tablehead{
\colhead{Region} & \colhead{Area} & \colhead{$N_{\rm H}$} &
\colhead{$\Gamma_b $} & \colhead{$\Gamma_s$} & \colhead{Intensity blank-sky} & \colhead{Intensity stowed} \\
\colhead{arcmin} & \colhead{$\rm arcmin^2$} & \colhead{$\rm 10^{22} cm^{-2}$} &
\nocolhead{Number} & \nocolhead{Number} & \colhead{$\rm 10^{-18} erg \, cm^{-2} \, s^{-1} \, arcsec^{-2}$} & \colhead{$\rm 10^{-18} erg \, cm^{-2} \, s^{-1} \, arcsec^{-2}$} 
}
\decimalcolnumbers
\startdata
0.17-1 & 2.04  & 0.7(frozen) & 1.60\pm0.07 & 1.65\pm0.08 & 12.9\pm1.60 & 12.4\pm1.65  \\       
1-2    & 6.03  & ~   & 1.59\pm0.07 & 1.59\pm0.07 & 7.28\pm0.88 & 7.43\pm0.90  \\
2-3    & 10.63 & ~.  & 1.53\pm0.11 & 1.58\pm0.12 & 3.12\pm0.62 & 3.08\pm0.64  \\
3-4    & 14.88 & ~   & 1.20\pm0.17 & 1.30\pm0.18 & 1.85\pm0.71 & 1.80\pm0.68  \\
4-5    & 17.80 & ~   & 1.31\pm0.25 & 1.30\pm0.22 & 1.15\pm0.60 & 1.33\pm0.54  \\
5-6    & 21.96 & ~   & 1.48\pm0.67 & 1.75\pm0.53 & 0.37\pm0.46 & 0.46\pm0.37  \\
\hline
0.17-1 & 0.55  & ~   & 1.33\pm0.19 & 1.35\pm0.19 & 14.5\pm5.76 & 14.3\pm5.63  \\       
1-2    & 1.61  & ~   & 1.68\pm0.14 & 1.75\pm0.14 & 7.69\pm1.75 & 7.83\pm1.68  \\
2-3    & 2.72  & ~   & 1.52\pm0.14 & 1.67\pm0.13 & 5.53\pm1.40 & 5.63\pm1.21  \\
3-4    & 3.66  & ~   & 1.64\pm0.12 & 1.74\pm0.13 & 5.24\pm1.06 & 4.83\pm0.98  \\
4-5    & 5.00  & ~   & 1.52\pm0.11 & 1.59\pm0.10 & 5.33\pm1.06 & 5.75\pm0.99  \\
5-6    & 5.74  & ~   & 1.49\pm0.10 & 1.50\pm0.10 & 5.50\pm1.02 & 5.79\pm1.06  \\
6-7    & 5.99  & ~   & 1.30\pm0.12 & 1.19\pm0.11 & 4.78\pm1.20 & 5.21\pm1.31  \\
\enddata
\tablecomments{(1) The region spectrum extracted from, the first line corresponds to the innermost annular region in Fig. \ref{fig:realpwn} (10 arcsec to 1 arcmin away from the pulsar). (2) Area of this region. (3) H I column density (frozen to 0.7). (4) Spectral POWER-LAW index analyzed with blank-sky background. (5) Spectral POWER-LAW index analyzed with stowed background. (6) Intensity of this region calculated from {\sc xspec} analyzed with blank-sky background. (7) Intensity of this region calculated from {\sc xspec} analyzed with stowed background. The six lines above are extracted from region {\sl a} while seven lines below are extracted from region {\sl b}.}
\end{deluxetable*}

Results of spectra fitting are shown in Table \ref{table:results}. Note that for the the blanksky analysis, the best-fit APEC parameters are kT $=1.05\pm0.11$ keV, norm $=(1.60\pm0.31)\times10^{-5}$. For stowed background, the parameters are kT $=0.43\pm0.12$ keV, norm $=(2.33\pm0.86)\times10^{-6}$ for unabsorbed APEC and kT $=3.78\pm0.26 $ keV, norm $=(2.04\pm0.07)\times10^{-5}$ for absorbed APEC respectively. Fitting results of blank-sky and stowed background are consistent with each other.
The intensity of region {\sl a} monotonically decreases with the increasing distance from the pulsar, which is inconsistent with a pure background. It is generally in line with the expectation in the pulsar halo scenario \citep{Liben22}, in which electrons are escaping from the PWN and diffusing towards large distance, forming an extended X-ray halo even beyond the edge of {\em Chandra} ACIS-I FoV. On the other hand, we can see that the intensity of region {\sl b} firstly decreases at the small radii, i.e., the first two annuli, which belongs to the compact PWN. There is no significant difference in these two annuli between region {\sl a} and region {\sl b}. However, when it goes farther away from pulsar, X-ray emission in region {\sl a} become fainter and hardly visible, while in region {\sl b} the intensity almost keeps constant to the edge of the FoV. 
It indicates different origin or transport mechanism of X-ray emitting electrons in these two regions. We did not extend the annuli to larger distance because it is too close to the edge of FoV where instrument response decreases and only covered by a few observations.  

The photon index profile is intriguing in both region {\sl a} and region {\sl b}. In region {\sl a}, the power-law spectra first become harder then become softer, while in region {\sl b}, the spectra first become softer and then become harder. Regardless of region {\sl a}, if the X-ray emitting electrons in region {\sl b} are transported away from the PWN, the spectra are expected to become softer if the radiation cooling is important, or keep constant if the cooling is negligible. The spectral hardening at large distance is not expected, unless certain reacceleration processes occur or it has another origin. However, in the latter two cases, a corresponding brightening in the intensity profile is supposed to appear at the location, which is unobserved. We note that the intensity in the local background region (the outermost annulus shown in Fig.~\ref{fig:realpwn}) reaches $8\times10^{-18} \, {\rm erg\, cm^{-2}\,s^{-1}\,arcsec^{-2}}$ . This value is actually comparable to the intensity of the obtained X-ray halo at large radii (i.e., beyond 3 arcmin from the pulsar). Therefore, the emission in the outermost annuli unlikely consists of a pure background. Indeed, HESS~J1809-193 is much more extended than the PWN and the FoV of {\rm Chandra}. In the IC interpretation, the background-extracting region must be filled with energetic electrons, probably escaping from the PWN. Hence the outermost annuli is likely a part of the X-ray halo and is not appropriate to be used as the background to reveal the properties of the X-ray halo itself. This may be the reason for the intriguing behaviour of the photon index profile.

\subsubsection{Spectral Analysis with Nearby Background}
As discussed above, the outermost annulus in Fig. \ref{fig:realpwn} is probably still dominated by the X-ray halo. Employing it as the background would lead to an overestimation of the background flux level and too soft the background spectrum.
While the obtained X-ray spectrum of the compact PWN may not be influenced significantly because of its high brightness, the spectral properties of the diffuse emission at large distance is supposed to be severely biased. We need to look for a more appropriate background region, and therefore searched in a larger area in the archival data of {\em Chandra} observations 
and chose several nearby observations as the background extracting regions (Fig. \ref{fig:rgb}). 

\begin{deluxetable*}{ccCCCCCCC}
\tablecaption{{\em Chandra} ACIS-I observations as nearby background candidate}\label{table:bkg obs}
\tablewidth{0pt}
\tablehead{
\colhead{Obsid} & \colhead{Target} & \colhead{Exposure} & \colhead{Area} 
& \colhead{kT (APEC)} & \colhead{Norm (APEC)} & \colhead{$N_{\rm H}$} & \colhead{$\Gamma$ (POWER-LAW)} & \colhead{Norm (POWER-LAW)} \\
\nocolhead{~} & \nocolhead{~} & \colhead{$\rm ks$} & \colhead{$\rm arcmin^2$} & \colhead{$\rm keV$} & \nocolhead{~} & \colhead{$\rm 10^{22} \, cm^{-3}$} & \nocolhead{~} & \nocolhead{~} 
}
\decimalcolnumbers
\startdata
18452 & W31 North           & 39.2 & 94.73 & 0.65\pm1.43 & (0.47\pm1.19)\times10^{-5} & 1.65 & 1.04\pm0.12 & (1.50\pm0.22)\times10^{-4} \\
18906 & M17 SW              & 59.3 & 29.34 & 0.87\pm0.30 & (9.38\pm4.52)\times10^{-6} & 1.23 & 0.59\pm0.19 & (2.89\pm0.72)\times10^{-5} \\
14662 & Suzaku J1811-1900   & 54.4 & 78.19 & 0.80\pm0.03 & (1.15\pm0.06)\times10^{-4} & 1.60 & 1.71\pm0.10 & (3.49\pm0.34)\times10^{-4} \\
16674 & W33                 & 38.2 & 65.78 & 0.76\pm0.04 & (1.10\pm0.07)\times10^{-4} & 1.46 & 0.98\pm0.08 & (1.66\pm0.17)\times10^{-5} \\
local & PSR J1809-1917      & 405  & 27.53 & 1.05\pm0.11 & (1.60\pm0.31)\times10^{-5} & 1.60 & 1.96\pm0.05 & (2.03\pm0.11)\times10^{-4} \\
\enddata
\tablecomments{(1) {\em Chandra} observation ID. (2) Target name of the observation. (3) Total exposure time. (4) Area of the region to extract spectra. (5) Plasma temperature of {\sc apec}. (6) Normalization of {\sc apec}. (7) Hydrogen column density. (8) Photon index of {\sc power-law}. (9) Normalization of {\sc power-law}. We also show the fitting results of local background in the bottom line for comparison.}
\end{deluxetable*}

\begin{deluxetable*}{CCCCCCC}
\tablecaption{PWN spectra fitting results with nearby background.}\label{table:spectra of nearby background}
\tablewidth{0pt}
\tablehead{
\colhead{Region} & \colhead{Area ({\sl a})} & \colhead{$\Gamma_{\sl a}$} & \colhead{Intensity ({\sl a})} & \colhead{Area({\sl b})}& \colhead{$\Gamma_{\sl b}$}  & \colhead{Intensity ({\sl b})}  \\
\colhead{arcmin} & \colhead{$\rm arcmin^2$} &  \nocolhead{Number} & \colhead{$\rm 10^{-18} erg \, cm^{-2} \, s^{-1} \, arcsec^{-2}$} & \colhead{$\rm arcmin^2$} & \nocolhead{Number} & \colhead{$\rm 10^{-18} erg \, cm^{-2} \, s^{-1} \, arcsec^{-2}$} 
}
\decimalcolnumbers
\startdata
0.17-1 & 2.04  & 1.77\pm0.07 & 14.1\pm1.50 & 0.55  & 1.54\pm0.17 & 15.7\pm4.84 \\       
1-2    & 6.03  & 1.84\pm0.06 & 8.58\pm0.73 & 1.61  & 1.93\pm0.12 & 8.84\pm1.43 \\
2-3    & 10.63 & 1.99\pm0.08 & 4.46\pm0.45 & 2.72  & 1.86\pm0.11 & 6.73\pm1.05 \\
3-4    & 14.88 & 2.09\pm0.09 & 3.10\pm0.33 & 3.66  & 1.99\pm0.10 & 6.31\pm0.81 \\
4-5    & 17.77 & 2.22\pm0.10 & 2.58\pm0.28 & 5.00  & 1.88\pm0.08 & 6.52\pm0.74 \\
5-6    & 21.96 & 2.35\pm0.12 & 1.94\pm0.23 & 5.74  & 1.85\pm0.08 & 6.71\pm0.76 \\
6-7    & 17.53 & 2.59\pm0.14 & 1.57\pm0.18 & 5.99  & 1.77\pm0.09 & 6.02\pm0.81 \\
\enddata
\tablecomments{(1) The region for spectrum extracted from. The first line corresponds to the innermost annulus in Fig. \ref{fig:realpwn} (10 arcsec to 1 arcmin away from the pulsar). (2) Area of the annulus in region {\sl a}. (3) Spectral power-law index of region a analyzed with blank-sky background. (4) Intensity of power-law spectrum of region {\sl a} calculated from {\sc xspec} analyzed with blank-sky background. (5) Area of the annulus in region {\sl b}. (6) Spectral power-law index of region {\sl b} analyzed with blanksky background. (7) Intensity of power-law spectrum of region {\sl b} calculated from {\sc xspec} analyzed with blank-sky background.}
\end{deluxetable*}

Detailed information on these observations and their fitting results are listed in Table \ref{table:bkg obs}. The four observations have similar Galactic latitudes to avoid systematical error of the Galactic diffuse X-ray background (Fig. \ref{fig:rgb}). Obsid 18452 is the closest observation to PSR~J1809-1917 while obsid 18906 is the farthest. Both of them are not associated with TeV emission. Obsid 14662 and 16674 are covered by extended emission of HESS J1809-193 and HESS J1813-178 separately. We analyzed their spectra of residual cosmic X-ray background of selected regions with {\sc apec+tbabs*powerlaw}. Fitting methods have been discussed in Section \ref{sec:3.1.1}. Plasma temperatures of these four spectra are similar while photon index varies. The spectrum of obsid 18096 is the hardest while  obsid 14662 is the softest. Photon index of obsid 18452 and 16674 are moderate and nearly the same. We ignored obsid 14662 because \cite{rangelov_multiwavelength_2014} suggested that probably there exists a TeV source powered by a dark accelerator in the northeastern outskirt of HESS J1809-193. Our X-ray blank-sky analysis shows that the fitting results of local background near PSR J1809-193 are similar to obsid 14662 field, which supports their argument. 

We joint-fit the other three observations to get an average background spectrum, which returns $ kT=0.67\pm0.08 \, $keV, $\rm \Gamma =0.92\pm0.07$. We take this as the residual of cosmic X-ray background and refit the PWN spectra and list the results in Table \ref{table:spectra of nearby background}. The resulting intensity profile shows the same trend with that obtained with the local background but the intensities are systematically higher in all annuli. On the other hand, the resulting photon index profile has been changed, showing a monotonic softening with increasing distance. Such a behaviour is consistent with the photon index profiles measured in some relatively young PWNe  \citep{Ge2021, hu_comprehensive_2022}, and is also consistent with the expectation of the X-ray halo produced by escaping electrons from the PWN. Strictly speaking, however, our result cannot demonstrate that the X-ray halo extends to the comparable size of HESS~J1809-193 due to the limited FoV of {\em Chandra}. In the future, advanced X-ray detectors with higher sensitivity and larger FoV may bring us the entire picture. 

On the other hand, the intensity profile and the photon index profile of region {\sl b} still show a more or less constant behaviour at large radii. It is clearly that the X-ray `tail' has a different origin or formation mechanism with respect to the X-ray halo in region {\sl a}.

\subsection{Modeling}
\subsubsection{Model Description}
\begin{figure}[htbp]
    \centering
    \includegraphics[width=0.5\textwidth]{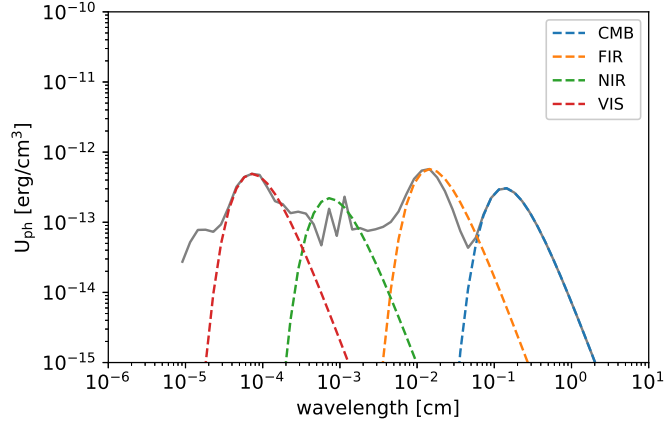}
    \caption{Radiation field that interacts with ultrarelativistic electrons by inverse Compton scattering. The black curve is the radiation field model from \cite{popescu_radiation_2017}, which we assumed four blackbody/graybody components to fit. The parameters are: CMB, T = 2.73 K, U$_{\rm ph}=4.2\times10^{-13}$ erg cm$^{-3}$; FIR, T = 31.67 K, U$_{\rm ph}=2.05\times10^{-12}$ erg cm$^{-3}$; NIR, T = 500 K, U$_{\rm ph}=4\times10^{-13}$ erg cm$^{-3}$; VIS, T = 5000 K, U$_{\rm ph}=1.5\times10^{-12}$erg cm$^{-3}$.}
    \label{fig:radiation_field}
\end{figure}

To test the IC interpretation of HESS~J1809-193, we need to check whether the the TeV emission can be explained together with the X-ray halo in the same framework.
We presume a simple one-zone model following \citet{liu_constraining_2019}, which is similar to that considered by \citet{di_mauro_evidences_2020}. In this model, we ignore the proper motion of PSR J1809-1917 and assume a homogeneous magnetic field as well as the diffusion coefficient in the space. We assume the injection electron spectrum $Q(E_e,t)$ to be a power-law with an exponential cutoff: 
\begin{equation}\label{eq:1}
   Q(E_e, t) = Q_0(t)\Big(\frac{E_e}{500 {\rm\ GeV}}\Big)^{-\alpha_e} e^{-E_e/E_{cut}}
\end{equation}
where $E_e$ is the energy of electrons, $\alpha_e$ is spectral index of electron spectrum, and $E_{cut}$ is the cutoff on electron energy. $Q_0(t)$ is the normalization factor that can be determined by
\begin{equation}
    \int_{E_{\rm min}}^{E_{\rm max}} E_e Q(E_e,t)dE_e = L(t)
\end{equation}
where $ E_{\rm min} =1$\,GeV and $ E_{\rm max} = $\,1 PeV. $L(t)$ is the spindown history of the pulsar which can be given by 
\begin{equation}
    L(t) = \eta L_i\Big(1+\frac{t}{\tau_0}\Big)^{-(n+1)/(n-1)}
\end{equation}
where $\eta$ is the efficiency of spin-down energy being converted to the injected electrons, $L_i$ is the initial spin-down luminosity, $n$ is the braking index which is set to be 3. The initial spin-down timescale of pulsar is given by
\begin{equation}
    \tau_0 = \frac{P_0}{2\dot{P_0}} = \tau_c - t_{\rm age}
\end{equation}
where $ t_{\rm age}$ is the age of pulsar that can be calculated by 
\begin{equation}
    t_{age} = \frac{P}{2\dot{P}}\Big[1 - \Big(\frac{P_0}{P}\Big)^2\Big] 
\end{equation}
where $P$ is rotation period of pulsar and $ P_0$ is its initial value, if $ P_0 << P$, it would reduce to 
\begin{equation}
    \tau_c = \frac{P}{2\dot{P}}
\end{equation}
where $ \tau_c$ is `characteristic age' which we use as pulsar's age. We also suppose an energy-dependent diffusion:
\begin{equation}\label{eq:7}
    D(E) = D_0\times\Big(\frac{E}{100 {\rm\ TeV}}\Big)^{\delta}
\end{equation}
where $D_0$ is the diffusion coefficient normailzed at 100\,TeV. The transport equation of electrons injected from a point source located at $r_s$ is
\begin{equation}
\begin{aligned}
\frac{\partial n_e(r, E_e, t)}{\partial t} &= D(E_e) \nabla_{r}^2 n_e + \frac{\partial[b(r,n_e,t)n_e]}{\partial E_e} + Q_e(E_e,t) \delta^3(r-r_s)
\end{aligned}
\end{equation}
where $n_e(r, E_e, t)$ is the differential electron density at time $t$ and position $r$, $D(E)$ is diffusion coefficient in Eq. (\ref{eq:7}), $Q_e$ is injected electron spectrum in Eq. (\ref{eq:1}), and $b(r,n_e,t)$ is the energy loss rate of electrons during the propagation, due to the synchrotron and IC radiation. Considering a homogeneous and constant magnetic field and background photon field, the energy loss can be estimated by \citep{cao_ultrahigh-energy_2021}
\begin{equation}\label{eq:9}
\begin{aligned}
b(E_e) &= -\frac{d E_e}{dt} = -\frac{4}{3}\sigma_Tc\Big(\frac{E_e}{m_ec^2}\Big)^2 \times \Bigg\{U_B+U_{ph,i}/\Big[1+\Big(\frac{2.82kT_iE_e}{m_e^2c^4}\Big)^{0.6}\Big]^{1.9/0.6}\Bigg\}
\end{aligned}
\end{equation}
where $\rm \sigma_T$ is Thmoson cross section, $\rm c$ is the speed of light, $\rm m_e$ is the mass of electron, $U_B = B^2/(8\pi)$ is energy density of magnetic field, $\rm k$ is Boltzmann constant, $\rm U_{ph}$ and $\rm T$ are energy density and corresponding temperature of photon field. For the background photon field that would interact with ultrarelativistic electrons through the IC scattering, we assume four blackbody/graybody components as cosmic microwave background (CMB), far-infrared radiation (FIR) field, near-infrared radiation (NIR) field, and visible light radiation (VIS) field around the pulsar as in Fig. \ref{fig:radiation_field}. The density of the background photon field is based on the model of \citet{popescu_radiation_2017}. 

The present-day $ (t = t_{\rm age})$ electron density with energy $E_e$ is given by   
\begin{equation}
\begin{aligned}
n_e(E_e,r) & =\int_{0}^{t_{age}} \frac{Q_e(E_g, t) dt}{(4\pi \lambda(E_e, t))^{3/2}} \times \exp\Big[-\frac{r^2}{4\lambda(E_e,t)}\Big] \frac{dE_g}{dE_e}
\end{aligned}
\end{equation}
where $\lambda(E_e,t)=\int_t^{t_{\rm age}}D(E_e'(t'))dt'$. Here, $E_e'(t')$ represents the energy of an electron is $E_e'$ at time $t'$ and evolves to $E_e$ at present $( t= t_{\rm age})$. And $E_g$ is the initial energy of an electron at its generation (injection) time $t$. The ratio $dE_g/dE_e$ can be calculated by tracing the energy evolution of electrons via Eq. (\ref{eq:9}) . Since the 70\% containment radius of HESS J1809-193 is 0.64\,degree assuming a Gaussian template \citep{2018A&A...612A...1H}, we model the profiles up to the maximum radius of $1^\circ$ which can cover most of the emission from HESS~J1809-193. 

\begin{figure*}[htbp]
    \centering
    \includegraphics[width=1.0\textwidth]{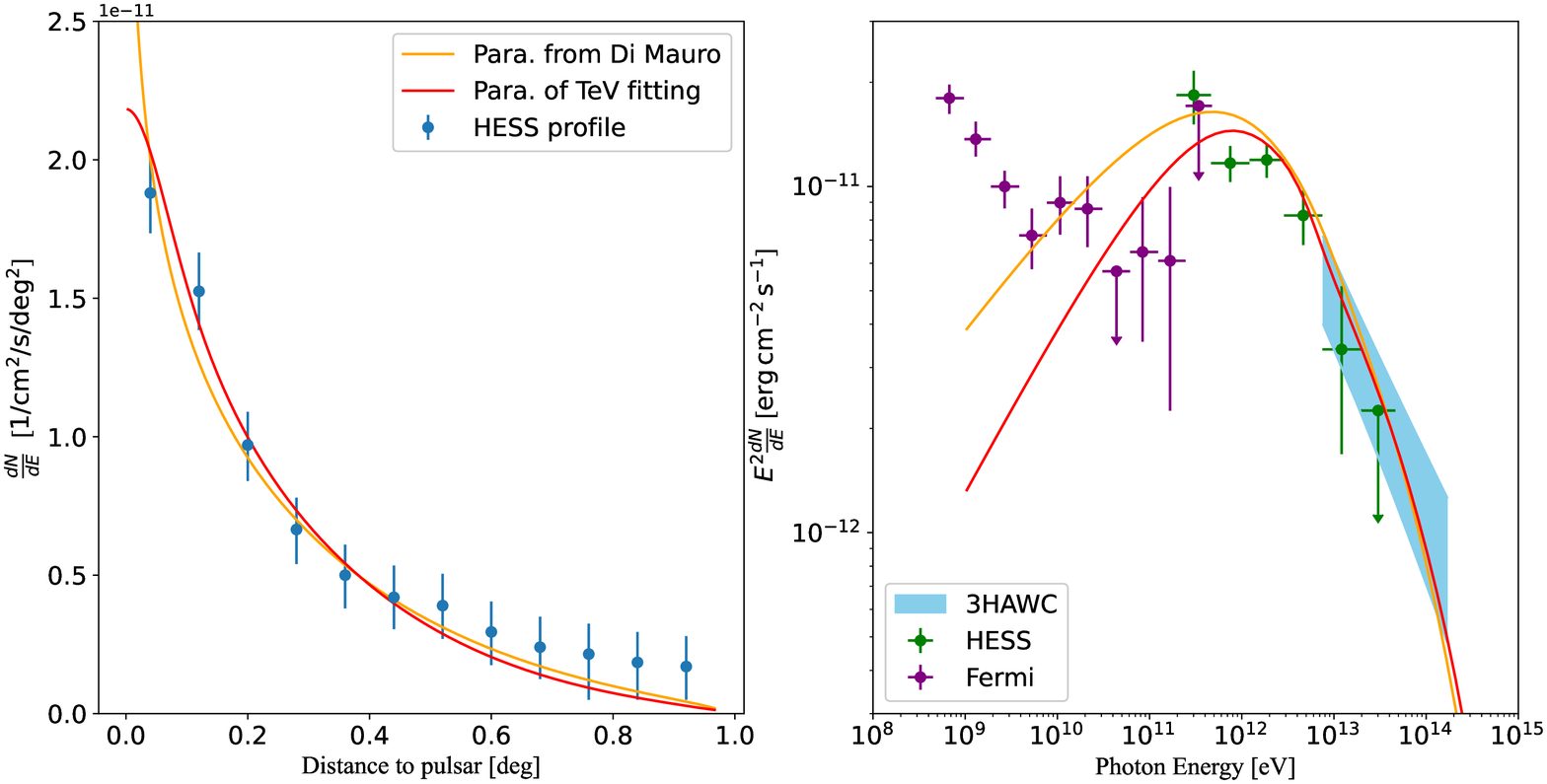}
    \caption{TeV radial profile (left) and SED (right) fitting with one-zone model. Orange, red lines are the prediction of our one-zone model with parameters from \cite{di_mauro_evidences_2020} and our best-fit parameters of TeV data. TeV profile data are taken from \cite{di_mauro_evidences_2020}(see also the footnote 6). Note that integral region to calculate SED is a circle with radius being 1 degree. The Fermi-LAT data is taken from \cite{araya_gev_2018} and HAWC data from \cite{albert_3hwc_2020}. The HESS data is taken from HESS Galactic plane survey (HGPS) source catalog.}
    \label{fig:tev}
\end{figure*}

\subsubsection{Pulsar Halo Scenario}
First, we checked our result following \cite{di_mauro_evidences_2020} who fit the TeV intensity profile of HESS J1809-193 in the pulsar halo scenario. We found that our model can reproduce the TeV intensity profile using their parameters\footnote{The spatially integrated TeV flux based on the 1D profile shown in \cite{di_mauro_evidences_2020} is twice the flux given by HGPS. Therefore we scaled the profile by a factor of 0.5. }. However, we found that the resulting flux with the same parameters overproduces the flux at $0.1-1\,$TeV by about 100\%. 
We therefore carry out a fitting to the TeV gamma-ray profile and the spectrum simultaneously with the Markov Chain Monte-Carlo method using the python pacakge {\em emcee}. When calculating the intensity profile, we take into account the influence of the limited angular resolution of HESS by convolving the theoretical intensity profile with the detector's point-spread-function. Considering a 68\% containment radius of the PSF of HESS to be $0.07\,\deg$, the intensity in the small radii becomes smoothed but it does not influence the profile at large radii. The result is shown in Figs.~\ref{fig:tev} and the best-fit parameters are listed in Table \ref{table:model fit}. We ignored the data below 100\,GeV which do not have the same trend as higher energy data, because they probably come from the gamma-ray pulsar. 

\begin{deluxetable*}{CCCCc}
\renewcommand{\arraystretch}{1.5}
\tablecaption{Best fit parameters using one-zone model\label{table:model fit}}
\tablewidth{0pt}
\tablehead{
\colhead{Parameter}  & \colhead{Best-fit of TeV} & \colhead{Best-fit of keV} & \colhead{Prior range of parameters}  & \colhead{Unit}
}
\decimalcolnumbers
\startdata
\eta         & 0.18^{+0.09}_{-0.06}    & 1.07^{+0.61}_{-0.32}\times10^{-2}  & [10^{-3},0]  & ~                \\
B            & 3.34^{+1.23}_{-1.18}    & 21.66^{+8.01}_{-6.99}  & [1,100] &        $\rm \mu Gauss$     \\
\alpha_e       & 2.06^{+0.13}_{-0.14}    & 1.29^{+0.46}_{-0.59}   & [1,3]   & ~                \\
D_0  & 1.18^{+0.67}_{-0.52}\times10^{28}    & 1.73^{+1.08}_{-0.78}\times10^{28} & [10^{25},10^{30}] & $\rm cm^2\, s^{-1}$ \\
\delta       & 0.65^{+0.23}_{-0.26}        & \sim 0                 & [0,1]   & ~                \\
\gamma_{cut} & 2.58^{+1.24}_{-1.22}\times10^9 & 1.11^{+0.49}_{-0.40}\times10^8   & [10^6,10^{10}]  & ~                \\
\enddata
\tablecomments{Best-fit parameters of our one-zone model. Definition of these parameters are: efficiency of transforming pulsar's spin-down luminosity to electrons in the wind; magnetic field strength; index of injected electron spectrum; diffusion coefficient; slope of diffusion coefficient; Lorentz factor of cutoff energy of injected electron spectrum.}
\end{deluxetable*}



\begin{figure*}[htbp]
    \centering
    \includegraphics[width=1.0\textwidth]{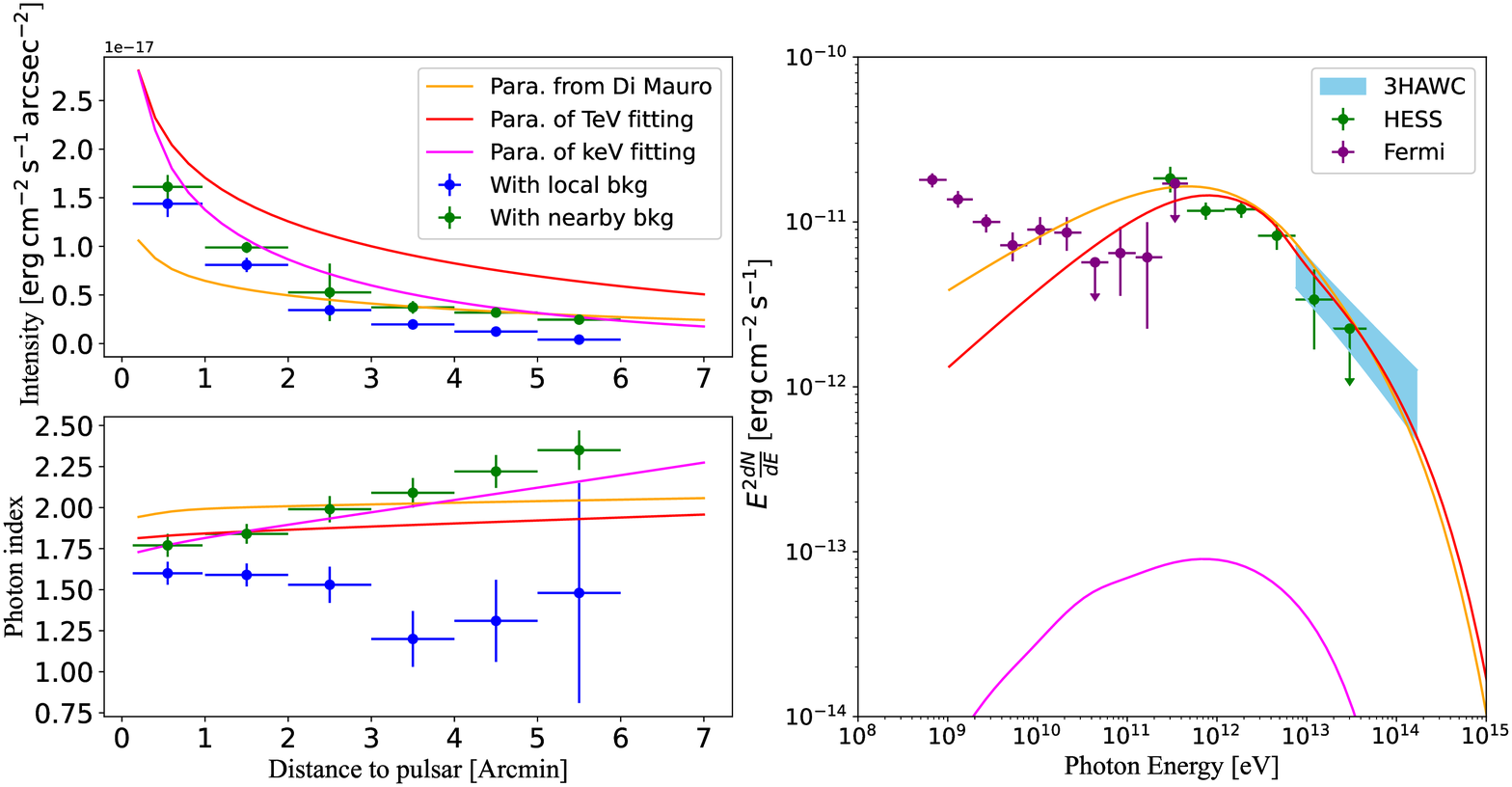}
    \caption{Left: keV radial profile of intensity and spectral index  from {\em Chandra} data. Orange and red lines are one-zone model prediction with the same parameters as in Fig. \ref{fig:tev}. Magenta lines are the best-fit to keV data with our one-zone model. Blue and green points are from region {\sl a} (Fig. \ref{fig:realpwn}) with local \& nearby blank-sky background subtracted separately. Right: Same as the right panel in Fig. \ref{fig:tev}, but with the prediction of the IC flux under the best-fit parameters for keV data added for comparison.}
    \label{fig:kev}
\end{figure*}

While our best-fit parameters successfully explain HESS data, the same parameters can not reproduce the X-ray emission as shown in Fig. \ref{fig:kev}. On one hand, the measured photon index of data softens much faster than the model prediction. On the other hand, the measured intensity is systematically lower than the model prediction. To reproduce the X-ray intensity profile and the photon index profile simultaneously, a stronger magnetic field of $\sim 21\,\mu$G is needed based on the same MCMC fitting procedure, so that the electrons can cool more efficiently. The best-fit parameters for the X-ray data are listed in Table ~\ref{table:model fit}. With this magnetic field, the IC radiation of electrons is suppressed. Integrating over a circular region around the pulsar with $1\deg$ radius, we found the predicted IC flux is lower than the measured one by more than one order of magnitude. In fact, in addition to the difference of the magnetic field required by the TeV-data fitting from that required by the keV-data fitting, some other parameters do not coincide in these two fittings neither.
For example, we limit the range of $\alpha_e$, the spectral index of injected electrons in the range of $1-3$. The keV-data fitting favors $\alpha_e\sim 1.2$ implying a hard electron spectrum, while the TeV-data fitting favors a soft electron spectrum with $\alpha_e \sim 2$. The inconsistencies between the keV data and TeV data may imply the existence of an additional TeV-emitting particle component, which can account for the flux measured by HESS and HAWC in the presence of the strong magnetic field required by the TeV data.

\subsection{Hadronic Scenario}
Pionic emission of high-energy protons from nearby supernova remnants have been suggested as the source of HESS~J1809-193, which could be the aforementioned additional particle component. To look into the hadronic scenario, we firstly assume a proton spectrum with power-law distribution followed by an exponential cutoff, according to the treatment of \citet{araya_gev_2018}. The semi-analytical model for gamma-ray from pion decay developed by \citep{kafexhiu_parametrization_2014} is employed in our calculation. We found, however, that the power-law spectrum cannot satisfactorily fit the gamma-ray SED, which is inconsistent with the conclusion of \citet{araya_gev_2018}. This is because \citet{araya_gev_2018} used an early version of HESS data \citep{aharonian_discovery_2007}, while \citet{2018A&A...612A...1H} announced that the updated SED of HESS~J1809-193 is different from the early version. In addition, \citet{araya_gev_2018} did not include the HAWC data. 

\begin{figure}[htbp]
    \centering
    \includegraphics[width=0.5\textwidth]{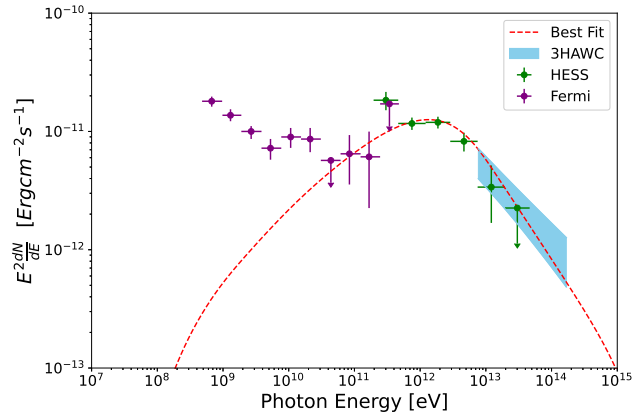}
    \caption{Hadronic model for HESS J1809-193. The data are the same as the right panel of  Fig. \ref{fig:tev}. The particle distribution is broken power-law (index 1.5 before break and 2.9 after break with break energy being 20 TeV).  }
    \label{fig:sed_hadronic}
\end{figure}

Then we employ a proton spectrum of the broken power-law distribution as follows:
\begin{equation}
f(E) = 
\begin{cases} 
A(E/E_b)^{-\alpha_1},  &  E < E_b \\
A(E/E_b)^{-\alpha_2},  &  E > E_b
\end{cases}
\end{equation}
where $E_b$ is the break energy. To fit the energy spectrum, a total energy of $W_{\rm p} = 2.6\times10^{50} (n/{1\,\rm cm^{-3} })^{-1} \, {\rm erg}$ is required, where $W_{\rm p}$ represents the total energy of hadrons and n is the density of the target hadrons. \cite{castelletti_unveiling_2016} proposed three molecular clouds near SNR G11.0-0.0 with an average proton density of $\sim 2.5\times10^3 \, {\rm cm}^{-3}$ can be the target for the hadronic interaction. Based on this, we obtain that the total energy of injected protons above 1 GeV is $1.1\times10^{47}\, {\rm erg}$, which is a tiny fraction of the expected CR energy budget of an SNR. Alternatively, ISM could also serve as the target for hadronic interactions if injected CRs are confined around the source (probably via the streaming instability, e.g., \citealt{2013ASSP...34..221G}). Indeed, for a typical hydrogen density in ISM, i.e., $n=1\,{\rm cm}^{-3}$, the required proton energy $W_{\rm p}=2.6\times 10^{50}\rm \,$erg is not extreme for an SNR. As shown in Fig.~\ref{fig:sed_hadronic}, the broken power-law proton spectrum can give a satisfactory fitting to the gamma-ray data above 0.1\,TeV. with $\alpha_1=1.5$, $\alpha_2=2.9$,  and $E_b=20\,$TeV. The break could be caused by the energy-dependent escape of CRs from SNRs so that a large fraction of low-energy CRs has not reached the molecular clouds \citep[e.g.,][]{Ohira11, Celli19, Bao21}. In addition, PWN or pulsars could also serve as the sources of these high-energy CR protons \citep{Cheng90, Blasi00, Amato03, the_lhaaso_collaboration_petaelectron_2021, Liang22}.

\section{Discussion and Conclusion}

In summary, we performed a detailed analysis of {\em Chandra} ACIS-I observations on the PSR J1809-1917 field, aiming to determine the association between its PWN and extended $\gamma $-ray source HESS J1809-193. Applying double background subtraction method, we investigated the morphology and X-ray spectra distribution of this PWN. Then we introduced a one-zone diffusion model to fit keV and TeV data.

On the smoothed, exposure-corrected image with quiescent particle background subtracted, the PWN presents a bright core surrounded by a compact nebula of about 2 arcmin. In the southeast region, there are some bright extended emission which may be a PWN tail formed by escaping particles. In the other directions, we found faint diffuse X-ray emission outside the compact nebula, showing a decreasing intensity profile with the increasing distance from the pulsar. It implies that the existence of an X-ray halo produced by electrons escaping the PWN. The halo may probably extend even beyond the {\em Chandra} FoV. As a result, the background region selected in the FoV could consist of the emission from the escaping the electrons, and thus the obtained flux and spectrum may be uncertain. We therefore selected background from {\em Chandra} ACIS-I observations on the regions outside the HESS~J1809-193 but not too far from the pulsar. The obtained X-ray intensity profile still presents a monotonic decline and, additionally, the spectrum appears a gradual softening with increasing distance from the pulsar. It is  consistent with our speculation of the existence of an extended X-ray halo around the compact PWN. Sensitive X-ray observations covering a larger FoV would be crucial to clarify this picture.

Next, we employed the one-zone model to fit the intensity profile and the photon index profile of the X-ray emission, assuming electrons are injected from the location of the pulsar and diffuse outwards. The best-fit parameters are, however, inconsistent with those for the HESS profile above 1 TeV. This is mainly due to a relative strong magnetic field, i.e., $21\,\mu$G, which is required to explain the rapid softening measured in the X-ray photon index profile. Such a strong magnetic field would severely suppress the IC radiation of the electrons and make it insufficient to reproduce the measured TeV flux of HESS~J1809-193. We further explored a hadronic origin of the $\gamma$-ray emission in this region and found it available under reasonable parameters, consistent with the results in some previous literature.

Technically, the X-ray data disfavors the IC interpretation in the framework of the simple one-zone diffusion model, but it does not necessarily exclude the leptonic origin of HESS~J1809-193. Taking into account more complicated processes, such as the joint play of advection and diffusion might explain the rapid softening of the X-ray emission in the extended halo with a weak magnetic field, which benefits the IC interpretation of the TeV flux. Besides, the diffusion coefficient and the magnetic field are assumed to be uniform in the one-zone diffusion model. With an inhomogeneous diffusion or magnetic field, the additional parameters would make the IC interpretation available. For example, \citet{2011ApJ...742...62V} introduced a multi-zone time-dependent model to simultaneously explain the X-ray and TeV data of HESS 1825-137. 
Therefore, the leptonic origin with a more sophisticated model may still remain a viable explanation to HESS~J1809-193. 

During the review process of this paper, HESS collaboration updated the observation of the HESS~J1809-193 region and resolved two components \citep{mohrmann_revisiting_2023, HESS_2023}: an extended, elliptic one described by an asymmetric Gaussian template surrounding a compact one fitted with a symmetric Gaussian template at the center. The extended component may be the TeV counterpart of the X-ray halo revealed in this study. 
A detailed theoretical modelling of the emission in both X-ray band and the TeV gamma-ray band is needed to elaborate the nature of this source.

\section*{Acknowledgement}
We thank the anonymous referee for the constructive report, and Hai-Ming Zhang, Ben Li and Kai Yan for the helpful discussions. This work is supported by the NSFC under grant No.U2031105.

\appendix
\section{Multiwavlength image of HESS J1809-193 and nearby field}
\begin{figure*}[h]
    \centering
    \includegraphics[width=1.0\textwidth]{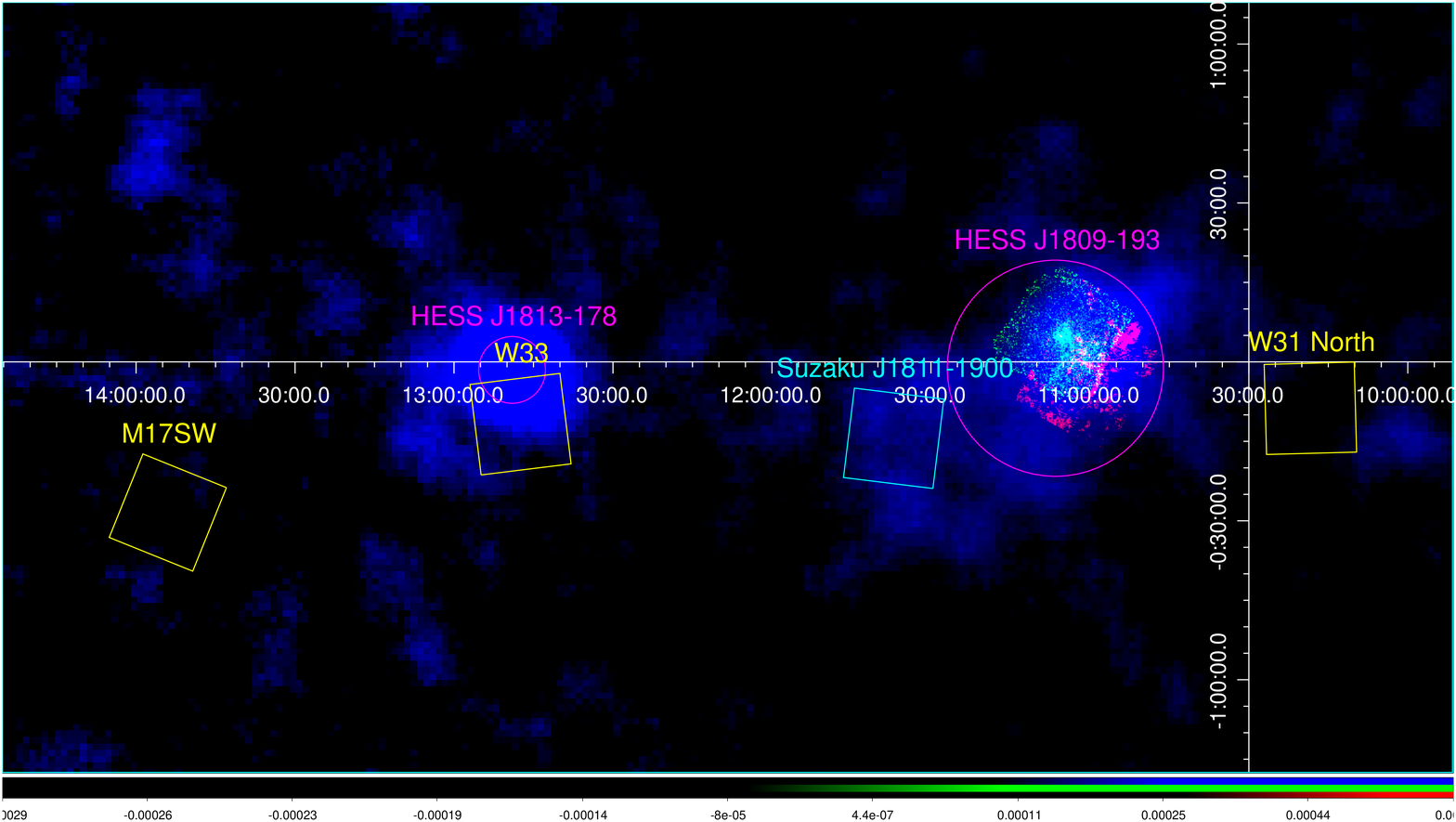}
    \caption{Multiwavlength image of HESS J1809-193 and its nearby region. Red: radio (JVLA, 1.4 GHZ, from \citealt{castelletti_unveiling_2016}); Green: X-ray ({\em Chandra} 0.5-7 keV, binned by a factor of 2 and smoothed with a 10$"$ Gaussian kernel); Blue: TeV (HESS Galactic Survey \citealt{2018A&A...612A...1H}).The squares show four nearby {\em Chandra} observations and corresponding target name we chose to be background candidates. Note that we excluded the observation to Suzaku J1811-1900 (cyan box) and combined the spectra of the other three observations (yellow boxes) to calculate the average background spectra. Magenta circles show 70\% containment radius of HESS sources.} 
    \label{fig:rgb}
\end{figure*}

Fig. \ref{fig:rgb} shows the multiwavlength image and the regions to extract spectra of nearby background regions.

\section{Fitting with different nearby backgrounds}

\begin{figure*}[h]
    \centering
    \includegraphics[width=1.0\textwidth]{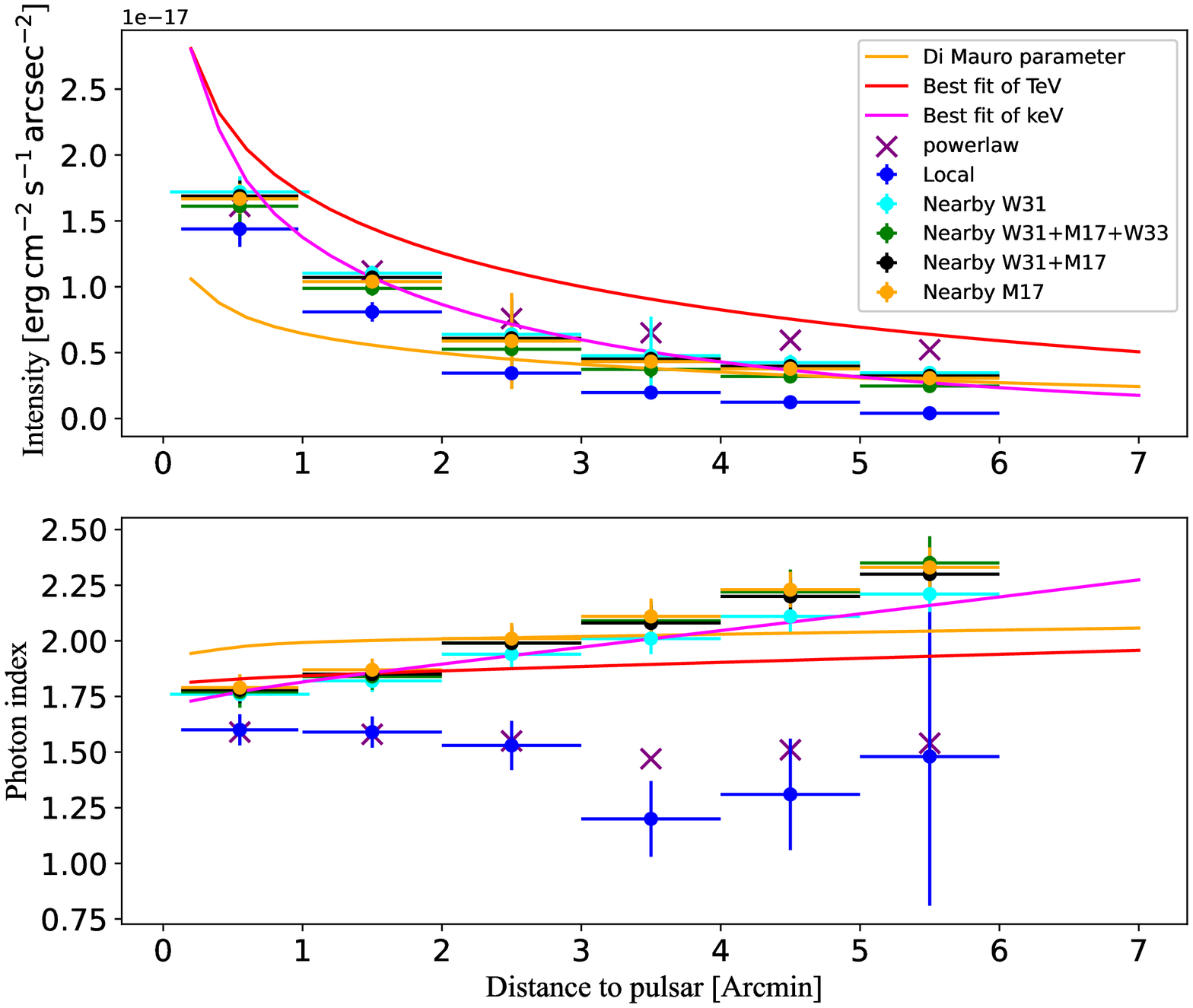}
    \caption{Radial profile of X-ray intensity(top) and photon index(bottom) with different backgrounds, similar to the left panel of Fig. \ref{fig:kev}. Cyan, orange and black points are calculated with W31, M17 and both as nearby backgrounds. We also show the results of ignoring residual sky background and fitting with pure absorbed power-law in purple X marker.}
    \label{fig:nbg_diff}
\end{figure*}
Fig. \ref{fig:nbg_diff} displays the fitting results with various choices of backgrounds. Our analysis reveals that the fitting results using different nearby backgrounds are consistent with each other, but are distinct from the results with using the local background or ignoring the residual sky background.

In addition to fitting with different choices of backgrounds, we also tried to fit the residual nearby backgrounds with different spectral models. The results are summarized in Table \ref{table: nbg fit}. No other models can fit with reasonable parameters or significantly improve the chi-square of the fitting, implying that our choice of background models in the main text is appropriate for the analysis.

\begin{deluxetable*}{cCCCCc}
\renewcommand{\arraystretch}{1.5}
\tablecaption{\textbf{Fit residual nearby background with various spectra models\label{table: nbg fit}}}
\tablewidth{0pt}
\tablehead{
\colhead{ Spectra model}  & \colhead{${\rm kT_1}$} & \colhead{${\rm kT_2}$} & \colhead{${\rm index_1}$}  & \colhead{${\rm index_2}$} & \colhead{$\chi^2$}
}
\decimalcolnumbers
\startdata
apec+tbabs*powerlaw   & 0.65\pm1.10   & ~  & 0.91\pm0.10  & ~  & 74.47/54            \\
tbabs*(apec+powerlaw) & 6.8\pm41.4^\star & ~ &0.70\pm1.35 & ~ & 75.93/54 \\
apec+tbabs*(apec+powerlaw) & 64.00\pm-1.00^\star & 0.04\pm0.02 & 0.93\pm0.10 & ~ &  71.11/52\\
 apec+powerlaw+tbabs*powerlaw & 0.009\pm-1^\star & ~ & 9.50\pm9.24^\star & 0.93\pm0.10 & 73.56/52 \\
apec+powerlaw+tbabs*(apec+powerlaw) & 0.001\pm10^{30\star} & 0.07\pm0.19 & 9.50\pm10^{15\star} & 0.93\pm0.10 & 71.65/50 \\
\enddata
\tablecomments{Best-fit parameters of residual nearby backgrounds (obsid 18452 and 18906) with various spectra models. The apec+tbabs*powerlaw model is the one used in the analysis shown in the main text. Unreasonable fitting parameters are marked with $^\star$ .}
\end{deluxetable*}


\clearpage
\bibliography{sample631}{}
\bibliographystyle{aasjournal}

\end{document}